\documentclass[fleqn,10pt]{wlscirep}
\usepackage[utf8]{inputenc}
\usepackage[T1]{fontenc}
\usepackage{ragged2e} 
\usepackage{amsmath}
\usepackage{amssymb}
\usepackage{graphicx}    % For including graphics
\usepackage{subcaption}
\usepackage{pifont}  % For the \ding command
\usepackage{xcolor}

\title{Uncovering the hidden core-periphery structure in hyperbolic networks}

\author[1,*,+]{Imran Ansari}
\author[1,*,+]{Pawanesh Yadav}
\author[1]{Niteesh Sahni}

\affil[1]{\ Department of Mathematics, Shiv Nadar Institution of Eminence Deemed to be University, Delhi-NCR-201314, India}

\affil[*]{\ ia717@snu.edu.in, py506@snu.edu.in}

\affil[+]{\ these authors contributed equally to this work}

\begin{abstract}

The hyperbolic network models exhibit very fundamental and essential features, like small-worldness, scale-freeness, high-clustering coefficient, and community structure. In this paper, we comprehensively explore the presence of an important feature, the core-periphery structure, in the hyperbolic network models, which is often exhibited by real-world networks. We focused on well-known hyperbolic models such as popularity-similarity optimization model (PSO) and $\mathbb{S}^{1}/\mathbb{H}^{2}$ models and studied core-periphery structures using a well-established method that is based on standard random walk Markov chain model. The observed core–periphery centralization values indicate that the core–periphery structure can be very pronounced under certain conditions. We also validate our findings by statistically testing for the significance of the observed core-periphery structure in the network geometry. This study extends network science and reveals core-periphery insights applicable to various domains, enhancing network performance and resiliency in transportation and information systems.

\end{abstract}
\begin{document}

\flushbottom
\maketitle
% * <john.hammersley@gmail.com> 2015-02-09T12:07:31.197Z:
%  Click the title above to edit the author information and abstract
%\thispagestyle{empty}
%\noindent Please note: Abbreviations should be introduced at the first mention in the main text – no abbreviations lists. Suggested structure of main text (not enforced) is provided below.

\section*{Introduction}

Complex networks are robust frameworks for analyzing and understanding complex systems in diverse domains. Over the years, the study of the complex network has revealed significant structure and interconnected patterns in the network science field. The field of application of complex networks is rapidly growing, ranging from understanding real-world systems to interdisciplinary fields, building new models and metrics, and addressing biological and social systems. The study focuses on revealing the statistical and topological properties that are fundamental to complex networks that represent complex systems \cite{albert2002statistical, newman2006structure,holme2012temporal}. In the past, researchers have studied real-world networks and established certain properties that characterize the complex networks, such as small-worldness \cite{milgram1967small}, a relatively high clustering coefficient \cite{watts1998collective}, heterogeneous degree distributions \cite{faloutsos1999power}, community structures \cite{fortunato2010community,fortunato2016community}, and the presence of core-periphery structure \cite{holme2005core}.

In the complex network, the interplay between nodes and edges reveal the significant patterns and structures. The most fascinating examples of such models, which are based on the degree and similarity of network nodes, are hyperbolic network models, such as the Popularity-Similarity Optimization (PSO) ~\cite{papadopoulos2012popularity} and the $\mathbb{S}^{1}/\mathbb{H}^{2}$ model~\cite{serrano2008self, garcia2019mercator} which are discussed in detail in the Methods section.
%%%%%%%%%%%%%%%%%% NEED TO WRITE THE PRECISE SUMMARY OF THE LEADING PAPER, Inherent community structure of the hyperbolic network.
Recently, in the year 2021, Bianka Kovács \& Gergely Palla \cite{kovacs2021inherent} have studied the hidden community structure of these hyperbolic network models comprehensively and discussed them for the various range of parameter settings. The authors generated the networks with a range of parameters such as popularity fading $\beta$ and temperature $T$ (average clustering coefficient) that varied in the plane $(0,1]\times[0,1)$ and the analogous parameters $(1/(\gamma-1),1/\alpha)$, where $\gamma$ is power law coefficient, $\alpha$ (average clustering coefficient) in the plane $(0,1)\times(0,1)$ for the PSO and $\mathbb{S}^{1}/\mathbb{H}^{2}$ model, respectively.  Then, they studied the community structure via various community detection algorithms. They claim that as the parameter settings go to the origin in both models, they yield the best community structure. Further, they analyzed the community structure as a function of a number of network nodes and claimed that community structure gets better for both networks for almost all the parameter settings. for the detailed study, please refer to the original papers \cite{kovacs2021inherent}.
%%%%%%%%%%%%%%%%%%%%%%%%%%%%%%%%%%%%%%%%%%%%%%%%%%%%%%%%%%%%%%%%   LITERATURE FOR THE CORE-PERIPHERY FOUNDATIONAL PAPERS, LEADING PAPERS OF CORE PERIPHERYNESS
Similarly to the community structure property of the network, core-periphery is another important aspect of network organization, where cohesive core nodes are surrounded by a sparse periphery. There has been many applications of the core-periphery structure, including social networks\cite{ Borgatti1999,Boyd2010, Rossa2013, holme2005core, Zhang2015, kojaku2017finding}, protein-protein interaction (PPI) networks \cite{Rossa2013, yang2014overlapping}, financial networks\cite{in2020formation, craig2014interbank}, transportation networks\cite{Rombach2017, Rossa2013,kojaku2017finding }, neural networks\cite{Rossa2013, tunc2015}. The core-periphery structure differs from community structures by highlighting densely connected core nodes that are also reasonably well connected to the periphery nodes. While homogeneous agents may not lead to unilaterally stable core-periphery networks, heterogeneity among agents can facilitate the formation of such structures. Various methods have been developed to detect and analyze core-periphery structures, showcasing their significance in understanding network dynamics and information flow across different domains\cite{Rossa2013,Boyd2010,holme2005core,surprise2018,Rombach2017,in2020formation,Borgatti1999}.

Network science has dedicated significant attention to unraveling the core-periphery structure, a pivotal mesoscale structure of networks in recent decades. The pioneering work by Borgetti and Everett\cite{Borgatti1999} laid the foundation for modeling core-periphery structures. They introduced algorithms for detecting core-periphery structures in weighted, undirected graphs, encompassing both discrete and continuous versions. Their discrete concept involves comparing a network to a block model comprising a fully connected core and a periphery devoid of internal edges but fully linked to the core. Their method aims to find a vector $\mathbf{C}$ of length $N$ whose entries can be either 1 or 0. The $i$th entry $C_i$ is equal to 1 if the corresponding node is assigned to the core, and equal to 0 if the corresponding node is assigned to the periphery. Let $C_{ij} = 1$ if $C_i = 1$ or $C_j = 1$, and let $C_{ij} = 0$ otherwise. Define $\rho_C$ as:

\begin{align*}
\rho_C &= \sum_{i,j} A_{ij} C_{ij},
\end{align*}

where the adjacency matrix element $A_{ij}$ represents the weight of the tie between the nodes $i$ and $j$ and equals 0 if the nodes $i$ and $j$ are not adjacent. This method of computing a discrete core-periphery structure seeks a value of $\rho_C$ that is high compared to the expected value of $\rho_C$ if $C$ is shuffled such that the number of 1 and 0 entries is preserved, but their order is randomized and in the continuous version of the core-periphery structure, wherein each node is assigned a "coreness" value denoted as $C_i$, with $C_{ij} = C_i \times C_j = a$.

Core-periphery structure, prevalent in various networks, exhibits diverse descriptions through different algorithms, such as k-cores decomposition and Borgatti-Everett's two-block model\cite{Borgatti1999}, leading to inconsistent interpretations and introducing a core-periphery typology and Bayesian stochastic block modeling aids in classifying networks, revealing a rich diversity of core-periphery structures critical for domain-specific analyses\cite{Gallagher2021}. Subsequently, following their seminal contributions, a plethora of methodologies have been developed by network scientists to detect core-periphery structures in various networks\cite{Borgatti1999,holme2005core,Tang2019,Rossa2013,Rombach2017,Zhang2015,Boyd2010,Lip2011,surprise2018}. These methodologies have been inspired notably by Borgetti and Everett's block-modeling approach, which involves partitioning networks into distinct components such as core-periphery or core-semiperiphery-periphery, or assigning continuous scores to individual nodes. The core-periphery structure is a versatile descriptor in various networks, but different algorithms can yield inconsistent descriptions, such as k-cores decomposition and the classic two-block model\cite{Gallagher2021}. In Ref.\cite{Gallagher2021}, Bayesian stochastic block modeling techniques are introduced to classify networks based on core-periphery typology, emphasizing the importance of acknowledging the diversity of core-periphery structures. By utilizing a connection density (CD) indicator and a region density (RD) curve, the paper\cite{Xiang2018} ranks nodes based on their connectivity to determine the presence of single CP structures, multiple CP structures, or community structures in a network. This approach enhances understanding of the relationships between different mesoscale structures in networks. In hyperbolic networks, core-periphery structure influences communication patterns, where peripheral nodes interact through core vertices, reflecting the tree-likeness and bending of shortest paths towards the core\cite{Alrasheed2015}. Birkan et al. \cite{Birkan2015} propose a unified approach for detecting and analyzing various mesoscale structures, enabling the examination of hybrid structures and statistical comparison. They illustrate its utility by analyzing the human brain network and uncovering dominant organizational structures (communities) and auxiliary features (core-periphery). The core-periphery (CP) structure, gaining prominence in complex networks, allows for discovering hidden network features. CP involves densely interconnected core nodes and sparsely connected periphery nodes, influencing various fields like economics and medicine. Despite its utility, comprehensive literature on CP detection problems and algorithms is lacking, highlighting its potential for further research\cite{Tang2019}.

%This paper provides a comprehensive and quantitative measure of core-periphery separation alongside an overall network centralization index\cite{Rossa2013}. To this end, we introduce a novel methodology that circumvents the need for artificial partitioning techniques and instead offers a nuanced portrayal of the network's core-periphery profile. This profile comprises a non-decreasing function denoted by $\alpha_1, \alpha_2, \ldots \alpha_n$, capturing the evolving core-periphery dynamics. In addition, we present a numerical indicator, termed the network centralization index $C$, which quantifies the extent of centralization within the network. This approach assigns a coreness value to each node and the network's overall centralization index, derived from the established random-walk Markov chain model, thereby facilitating a more nuanced understanding of network structure and dynamics. 

 In this paper, we generate random graphs using the PSO and $\mathbb{S}^{1}/\mathbb{H}^{2}$ models across a comprehensive range of parameter settings to examine their core-periphery structure. Employing the well-established core-periphery structure search algorithm by Rossa et al.\cite{Rossa2013} using the core-periphery centralization index, we thoroughly investigate these structures. Furthermore, we validate our findings through rigorous statistical testing to assess the significance of the observed core-periphery structures in these hyperbolic models. We focus on extending the modelling capabilities of hyperbolic network models, particularly within the context of core-periphery structures. Although these models, as discussed in the literature \cite{garcia2019mercator,papadopoulos2014network}, capture essential network properties, their applicability to core-periphery structures remains under-explored. Our study involves extensive simulations, parameter space exploration, and the use of advanced core-periphery detection algorithms to identify core-periphery regions. The results have implications for modelling real-world networks with core-periphery organizations and contribute to the broader understanding of network science. This research represents a concerted effort to advance our knowledge of core-periphery structures within hyperbolic network models, particularly the PSO model and the $\mathbb{S}^{1}/\mathbb{H}^{2}$ model. By systematically investigating their ability to capture core-periphery patterns, we aim to demonstrate the versatility of hyperbolic models in representing diverse network structures.

\section*{Results}

This manuscript comprehensively explores the core-periphery structure of the random graphs constructed using the two hyperbolic network models PSO and \(\mathbb{S}^1/\mathbb{H}^2\) across the various parameter configurations. These networks were then fed into the core-periphery detection model. Our findings highlight that hyperbolic random graphs possess a significant core-periphery structure within the wide range of parameter space.

\begin{figure}[htbp]
    \centering
    \begin{subfigure}[b]{0.48\textwidth}
        \centering
        \includegraphics[width=\textwidth]{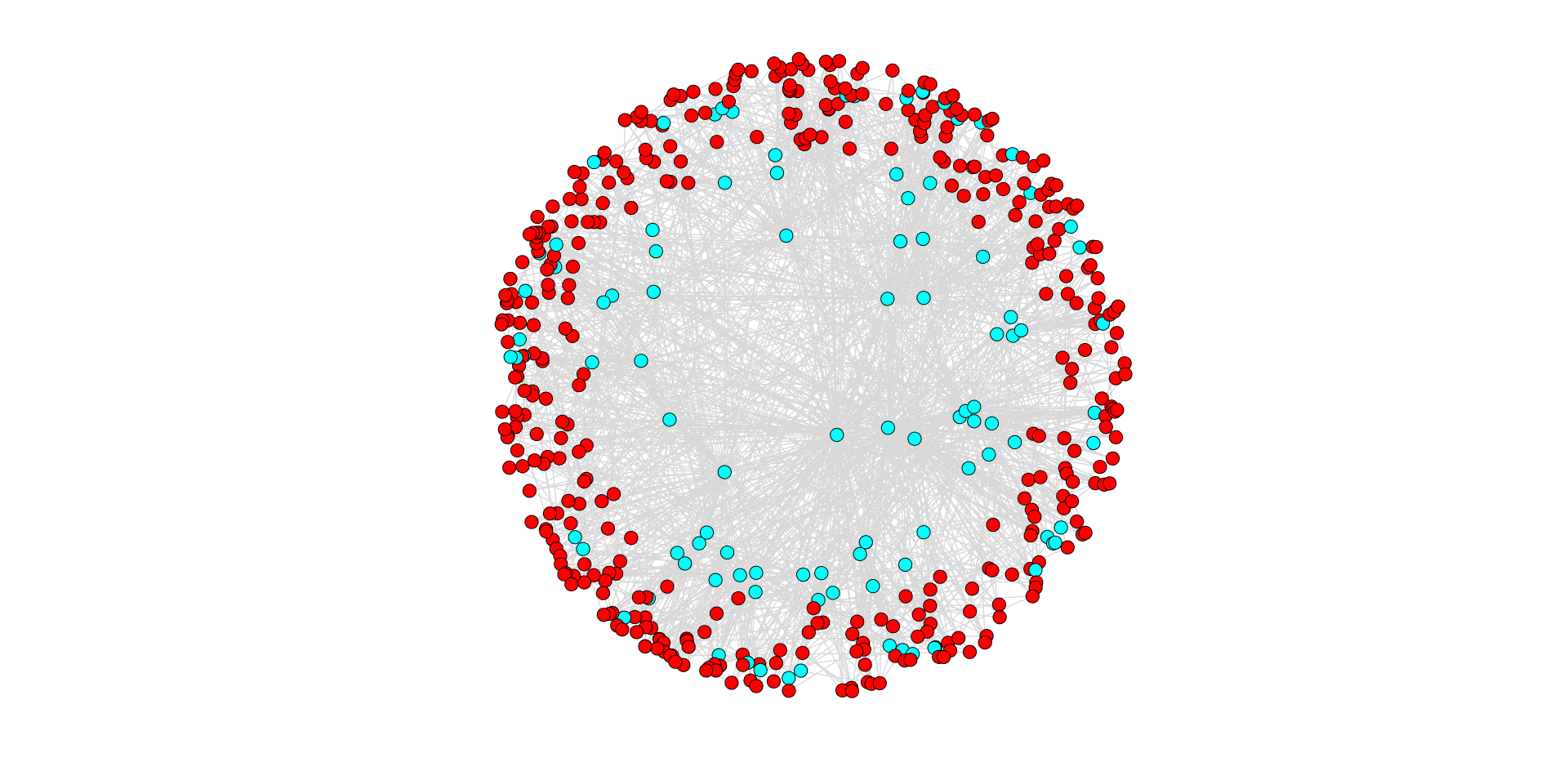}
        \caption{PSO Network }
        \label{fig:sub1}
    \end{subfigure}
    \hfill
    \begin{subfigure}[b]{0.48\textwidth}
        \centering
        \includegraphics[width=\textwidth]{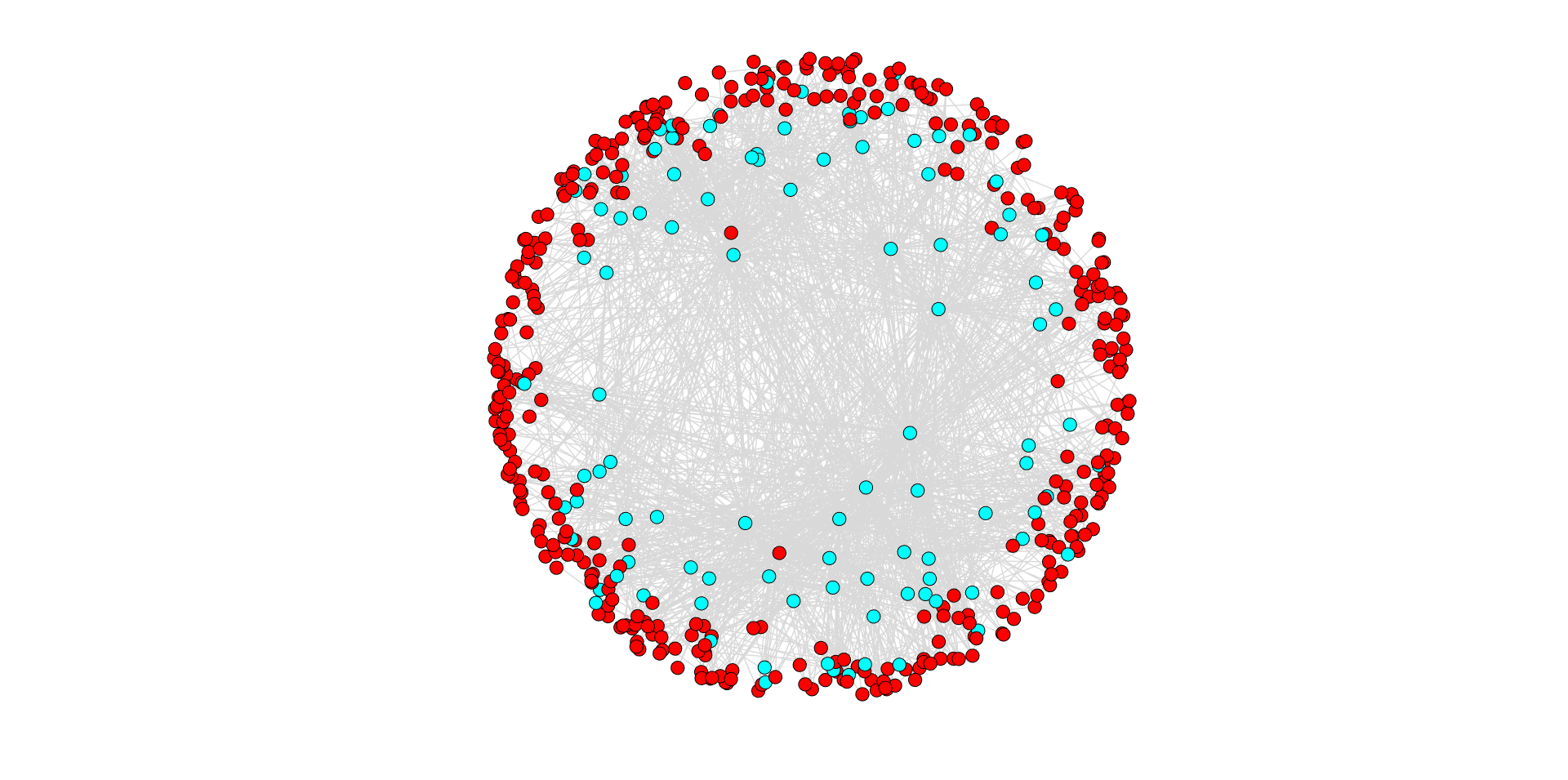}
        \caption{ \(\mathbb{S}^1/\mathbb{H}^2\) Network}
        \label{fig:sub2}
    \end{subfigure}
    \caption{Core-periphery visualisation in hyperbolic networks. (a) Core (cyan) and periphery (red) obtained in a network with $N = 500$ number of nodes, generated by the PSO model with parameters $m = 10$ (corresponding to $<k> = 20$), $\beta = 0.8$ (corresponding to $\gamma = 2.25$) and $T = 0.8$. The layout shows the network in the native disk representation of the two dimensional hyperbolic space of curvature $K =-1$, with the nodes arranged according to their coordinates assigned during the network generation process. (b)  Core (cyan) and periphery (red) obtained in a network generated by the \(\mathbb{S}^1/\mathbb{H}^2\) model with parameters $N = 500$, $<k> = 20$, $\gamma = 2.25$ and $\alpha = 1.125$, shown in the native disk representation of the hyperbolic plane of curvature $K =-1$}
    \label{fig:main}
\end{figure}

Fig.~\ref{fig:main}, represents an example of the core-periphery structure in which the core and periphery are identified by the Rossa algorithm within the network of size $N = 500$. We visualized the \(\mathbb{S}^1/\mathbb{H}^2\) and PSO networks in the two-dimensional native hyperbolic disk layout in Figures ~\ref{fig:main} (a) and ~\ref{fig:main} (b), respectively. In both networks, we visualized the top 100 as the core nodes (cyan) and the remaining 400 as the peripheral nodes (red). All core nodes are placed near the origin of the disk nodes, whereas the periphery is placed near the disk's boundary.

In addition, we computed the core-periphery (cp-centralization) values for each network using the Rossa algorithm. To do so, we keep the parameter configurations described in the article \cite{kovacs2021inherent}. According to this referenced paper, for the PSO model, the two parameters temperature $T$ and the popularity fading $\beta$ are equidistantly sampled in 10 data points between 0 and 1. Thus, we get 100 pairs of parameters $(T, \beta)$ in the parameter plane $T - \beta$. For each parameter setting, we generated 100 networks. On the other hand, To keep a direct one-to-one comparison of the parameters with the PSO model, in the model \(\mathbb{S}^1/\mathbb{H}^2\),  we replace the actual parameters $\alpha$ and $\gamma$ with the $1/\alpha$, $1/(\gamma-1)$ (analogous to temperature $T$ and popularity fading parameter $\beta$ in the PSO model), respectively. We sample the 90 pairs of combinations in the parameter space of $(1/\alpha) - 1/(\gamma - 1)$. This setting is based on finite values of $\alpha$ and $\gamma > 2$, so we excluded $T = 1/\alpha = 0$ and $\beta = 1/(\gamma - 1) = 1$ points from the analysis. Similarly, for the PSO model, we generated 100 networks for each combination of parameters.

\begin{figure}[htbp]
    \centering
    \includegraphics[width=1.0\textwidth]{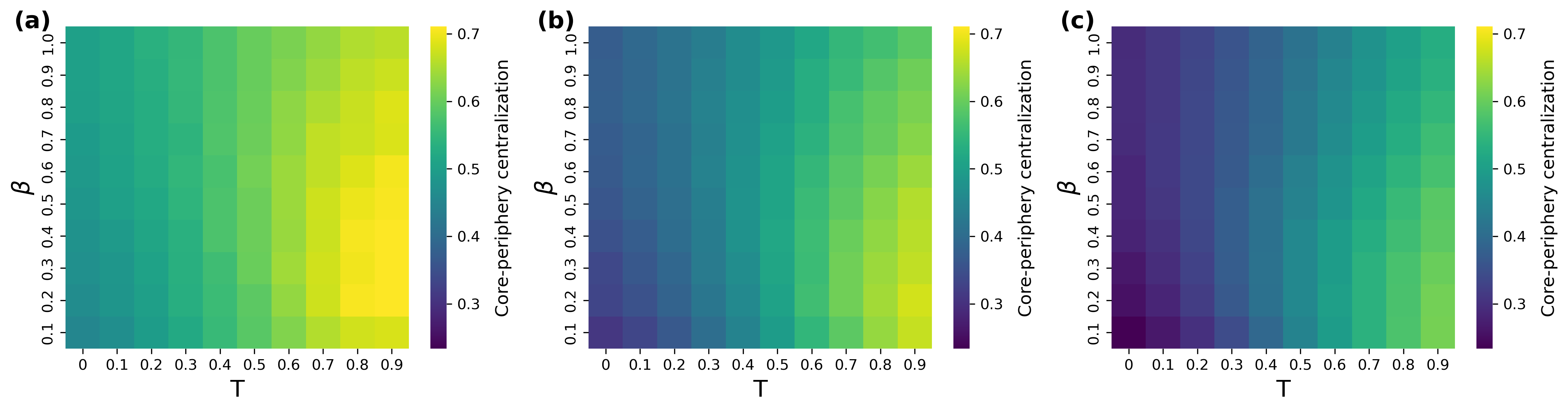}
    \caption{Core-periphery centralization in the PSO model. We show the core-periphery centralization $C$ as a function of the model parameters $T$ and $\beta$ for networks of size $N = 100$ and the expected average degree: (a). $\langle k \rangle = 4$, (b). $\langle k \rangle = 10$, and (c). $\langle k \rangle = 20$.}
    \label{fig:heatmaps1}
\end{figure}

\begin{figure}[htbp]
    \centering
    \includegraphics[width=1.0\textwidth]{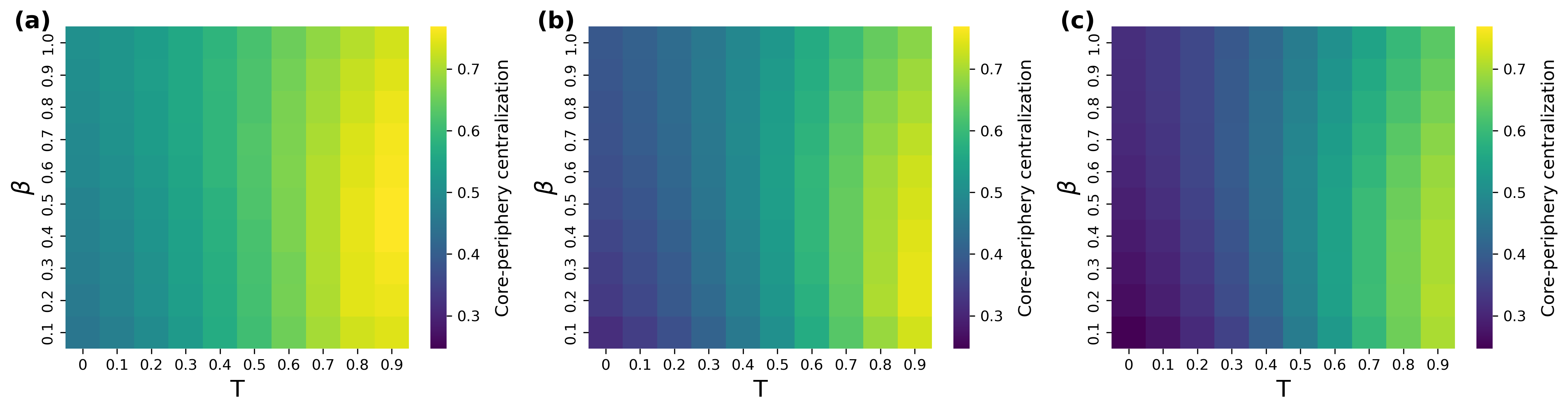}
    \caption{Core-periphery centralization in the PSO model. We show the core-periphery centralization $C$ as a function of the model parameters $T$ and $\beta$ for networks of size $N = 500$ and the expected average degree: (a). $\langle k \rangle = 4$, (b). $\langle k \rangle = 10$, and (c). $\langle k \rangle = 20$.}
    \label{fig:heatmaps2}
\end{figure}

\begin{figure}[htbp]
    \centering
    \includegraphics[width=1.0\textwidth]{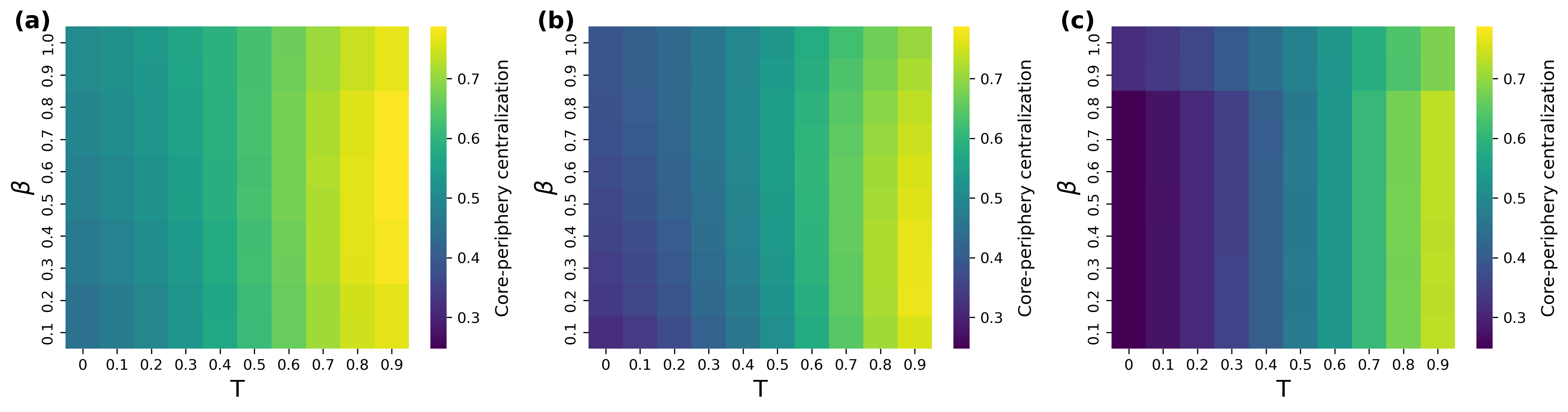}
    \caption{Core-periphery centralization in the PSO model. We show the core-periphery centralization $C$ as a function of the model parameters $T$ and $\beta$ for networks of size $N = 1000$ and the expected average degree: (a). $\langle k \rangle = 4$, (b). $\langle k \rangle = 10$, and (c). $\langle k \rangle = 20$.}
    \label{fig:heatmaps3}
\end{figure}

Next, we present the heat maps of the corresponding core-periphery (cp-centralization) $C$ given in Eq. ~\ref{eq:six} as a function of the model parameters. In Fig. ~\ref{fig:heatmaps1} (a), ~\ref{fig:heatmaps1} (b) and ~\ref{fig:heatmaps1} (c), we show the results for the PSO network of size $N=100$ and the expected average degree $<k> = 4$, $<k> = 10$ and $<k> = 20$,  respectively. Here, the cp-centralization value is averaged over 100 generated networks corresponding to a parameter $(T, \beta)$.  According to Fig. ~\ref{fig:heatmaps1} (a), for considerably higher temperature $T\geq 0.6$ and for any $\beta$, cp-centralization is in the range 0.60 to 0.72, indicating the presence of a strong and significant core-periphery structure in the sense of Rossa et al. Consequently, in Figs. ~\ref{fig:heatmaps1} (b) and \ref{fig:heatmaps1} (c), as we increase the expected average degree as $<k>=10$ and $<k> = 20$, the cp-centralization score starts to fall compared to the average degree $<k>=4$ and the high values of cp-centralization tend to cluster at $T=0.9$ and $\beta=0.1$ in Fig. \ref{fig:heatmaps1} (c). Furthermore, we increase the number of nodes $N = 500$ and $N = 1000$ and keep the expected average degrees similar to $N =100$ in Figs. ~\ref{fig:heatmaps2} and ~\ref{fig:heatmaps3}, respectively.  Thus, we observe in all three networks of size 100, 500 and 1000 that the heat maps corresponding to the expected degree $<k>=4$ have higher cp-centralization values than those corresponding to $<k>=10$ and $<k>=20$ in figures ~\ref{fig:heatmaps1}, ~\ref{fig:heatmaps2} and ~\ref{fig:heatmaps3} for all the parameter range of $(T, \beta)$.

\begin{figure}[htbp]
    \centering
    \includegraphics[width=1.0\textwidth]{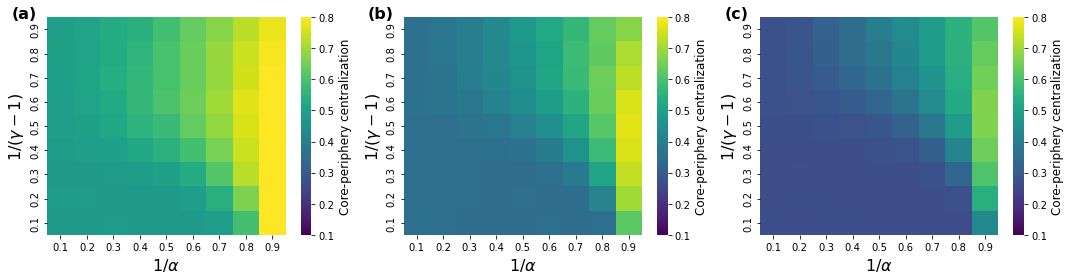}
   \caption{Core-periphery centralization in the \(\mathbb{S}^1/\mathbb{H}^2\) model. We show the core-periphery centralization $C$ as a function of the model parameters $1/\alpha$ and $1/(\gamma-1)$ for networks of size $N = 100$ and the expected average degree: (a). $\langle k \rangle = 4$, (b). $\langle k \rangle = 10$, and (c). $\langle k \rangle = 20$.}
    \label{fig:heatmaps4}
\end{figure}

\begin{figure}[htbp]
    \centering
    \includegraphics[width=1.0\textwidth]{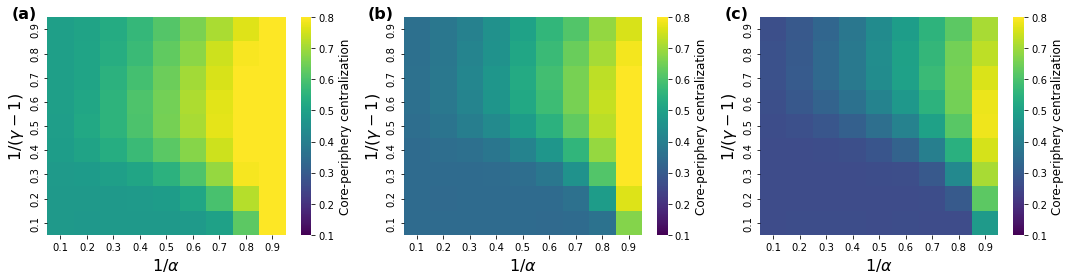}
    \caption{Core-periphery centralization in the \(\mathbb{S}^1/\mathbb{H}^2\) model. We show the core-periphery centralization $C$ as a function of the model parameters $1/\alpha$ and $1/(\gamma-1)$ for networks of size $N = 500$ and the expected average degree: (a). $\langle k \rangle = 4$, (b). $\langle k \rangle = 10$, and (c). $\langle k \rangle = 20$.}
    \label{fig:heatmaps5}
\end{figure}

\begin{figure}[htbp]
    \centering
    \includegraphics[width=1.0\textwidth]{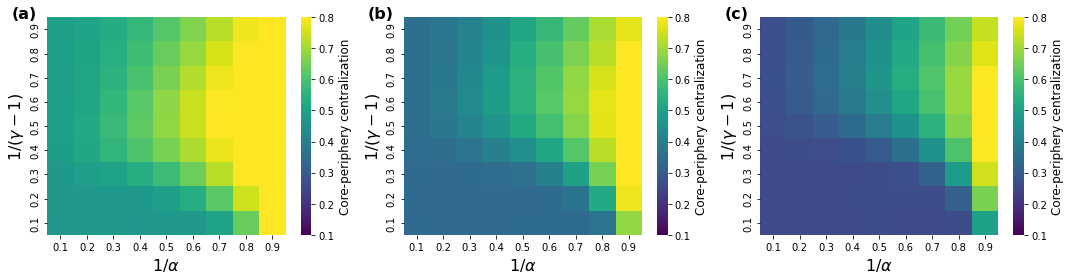}
    \caption{Core-periphery centralization in the \(\mathbb{S}^1/\mathbb{H}^2\) model. We show the core-periphery centralization $C$ as a function of the model parameters $1/\alpha$ and $1/(\gamma-1)$ for networks of size $N = 1000$ and the expected average degree: (a). $\langle k \rangle = 2$, (b). $\langle k \rangle = 10$, and (c). $\langle k \rangle = 20$.}
    \label{fig:heatmaps6}
\end{figure}

Similar, patterns has been observed in the \(\mathbb{S}^1\) network model with the same setting of the parameters of $T$ and $\beta$ is provided in Figures ~\ref{fig:heatmaps4},  ~\ref{fig:heatmaps5} and ~\ref{fig:heatmaps6}.

\section*{Discussion}

\begin{justify}

This section sheds light on the core-periphery structure problem in both the PSO and \(\mathbb{S}^1/\mathbb{H}^2\) models through in-depth testing. The networks produced by these approaches exhibit strong core-periphery structures for a broad range of model parameters despite the absence of intentional core-periphery structure. This is demonstrated by the high centralization values of the core-periphery measured on the results of the core-periphery structure algorithm, as provided by Rossa et al., described in Section 3.

The parameter plane in which we observed the behavior of the core-periphery centralization (cp-centralization) of the core-periphery corresponded to the $(T, \beta) \in [0, 1) \times (0, 1]$ plane in the PSO model and the analogous $(1/\alpha, 1/(\gamma - 1)) \in (0, 1) \times (0, 1)$ plane in the \(\mathbb{S}^1/\mathbb{H}^2\) model. The intuitive meaning of these parameters can be summarized as follows: the average clustering coefficient of the generated network is controlled by temperature $T$ and its equivalent $1/\alpha$. In contrast, the power-law decay exponent $\gamma$ of the degree distribution is controlled by the popularity fading parameter $\beta$ in the case of the PSO model according to the formula $\gamma = 1 + 1/\beta$ and is itself a parameter of the \(\mathbb{S}^1/\mathbb{H}^2\) model. Our findings indicate that the behavior of the cp-centralization for both hyperbolic models, PSO and \(\mathbb{S}^1/\mathbb{H}^2\), is comparable when these parameters are changed. The cp-centralization increases with an increase in the average clustering coefficient $T$ (or $1/\alpha$), and this centralization increases again with an increase in $\beta$ (or $1/(\gamma - 1)$). 

However, we find that the cp-centralization is not at all linearly dependent on the model parameters; rather, the lowest centralization values of the core-periphery centralization are produced when we consider the parameter settings close to the origin ($T \to 0$, $\beta \to 0$ in the PSO model and $1/\alpha \to 0$, $1/(\gamma - 1) \to 0$ in the \(\mathbb{S}^1/\mathbb{H}^2\) model); however, centralization continues to increase if these parameter move away from the origin. 

%It is worth noting that the associated networks do not resemble real networks without scale and are homogeneous in degree.

This regime exists in the parameter space, which appears to be consistent with the small-world transition found in \cite{garcia2018multiscale} by the renormalization group approach; that is, where the networks are highly local, and the core-periphery structures are strongest, the small-world property vanishes under renormalization. However, when $T$ increases (or $1/\alpha$ increases, controlling the clustering coefficient), centralization increases after some threshold for some range. For example, the centralization averaged across 100 networks can still reach $\mathrm{C} \approx 0.7$ in the PSO model and $\mathrm{C} \approx 0.75$ in the \(\mathbb{S}^1/\mathbb{H}^2\) model after a threshold $T = 0.6$, equal to $\alpha \approx 1.66$.

In other words, when setting the degree decay exponent to moderate values often observed in real systems, with the help of $\beta$ or by directly tuning $\gamma$, the network obtained with the studied model still has a core-periphery structure if the other parameters ($T$ or $1/\alpha$) are not pushed to extremely low values, which means that the clustering coefficient is reduced to extremely low values.

On the one hand, the regime where \( C \) declines to lower values is where \( T \rightarrow 0 \), corresponding to networks with clustering coefficients near to zero, and where \( \beta \rightarrow 0 \), corresponding to extremely fat-tailed degree distributions. Therefore, it could be preferable to select the models in \cite{zuev2015emergence,muscoloni2018nonuniform,alessandro2018leveraging,garcia2018soft} if one would like to create scale-free hyperbolic networks with core-periphery structures and a degree decay exponent that must be quite big. However, the examined "traditional" hyperbolic models appear to provide a robust enough core-periphery structure, except the previously described extreme regimes, to be regarded as a basic model for the apparent modular structures frequently found in real systems.

Why do the observed core-periphery structures arise without any obvious core-periphery formation mechanism built into the studied models? In short, the same model properties that allow the development of a small clustering coefficient in random graphs generated at the level of nodes also make the emergence of core-periphery structures possible at a slightly lower scale. Core-periphery structures are local structures in the sense that core nodes connect to each other with a larger link density than those at the periphery.

Our opinion is that the primary factor in the formation of core-periphery structures in the models under study is that, as demonstrated by the distance formula in Eq. ~\ref{eq:second}, it is far simpler for a node that has recently appeared at the periphery to connect radially than nodes with similarly large radial coordinates because of the hyperbolicity of the native disk. With enough angular separation, the previously arrived nodes positioned at smaller radii can develop into unique and appealing cores to which the new nodes can connect with minimal interference between the various angular regions. In the PSO model, the inner nodes must be forced outward for there to be a sufficient distance between them.

\section*{Statistical Significance of the Results}

To assess the statistical significance of the observed cp-centralization value \( C \), we employ a rigorous computational approach to calculate the \( p \)-value. This involves generating 100 randomized networks for each original network, ensuring that these randomized networks preserve the same degree distribution as the original\cite{hakimi1962realizability, kleitman1973algorithms}. For each of these randomized networks, we compute the cp-centralization, denoted as \( C_{i}^{\text{rand}} \) for \( i = 1, 2, \ldots, 100 \). Each cp-centralization values was obtained by averaging the cp-centralization of 100 networks corresponding to given parameters of the PSO and \(\mathbb{S}^1/\mathbb{H}^2\) models. The \( p \)-value is determined by the proportion of randomized networks with a cp-centralization value greater than the observed \( C \), calculated as \( p = \frac{\#\{i : C_{i}^{\text{rand}} > C\}}{100} \). Based on the calculated \( p \)-values, we test the following hypotheses: 

\textbf{Null Hypothesis} (\( H_0 \)): The observed cp-centralization \( C \) is not statistically significant and could have arisen by chance. 

\textbf{Alternative Hypothesis} (\( H_a \)): The observed cp-centralization \( C \) is statistically significant, suggesting it is unlikely to have occurred randomly.

A small \( p \)-value (typically less than 0.05 or 0.1) leads us to reject the null hypothesis \( H_0 \) and accept the alternative hypothesis \( H_a \), thus concluding that the observed cp-centralization \( C \) is significant and not a result of random variation.

The statistical significance of the core-periphery structure for the PSO networks corresponding to the parameters \( T \) and \( \beta \) was evaluated using \( p \)-values. For the networks of size \( N = 100 \), we observed that for \(\langle k \rangle = 4\), all parameters(100\%) exhibited significant core-periphery structures at the 5\% significance level. As the expected average degree increased to \(\langle k \rangle = 10\), the proportion of significant parameters decreased to 75\% at the 5\% level and 88\% at the 10\% level. For \(\langle k \rangle = 20\), only 46\% of the parameters were significant at the 5\% level and 57\% at the 10\% level. Similar trends were observed for networks of size \( N = 500 \) and \( N = 1000 \), with higher average degrees resulting in a lower proportion of significant parameters. These results indicate that increasing network density diminishes the statistical significance of the PSO model's core-periphery structure.

On the other hand \(\mathbb{S}^1/\mathbb{H}^2\) networks of size \( N = 100 \) with average degree 4,10 and 20, corresponding to parameters \(1/\alpha\) and \(1/(\gamma-1)\) also exhibited significant core-periphery structures at the 5\% significance level for \(\langle k \rangle = 4\) and \(\langle k \rangle = 10\). However, for \(\langle k \rangle = 20\), the proportion of significant parameters decreased to 63.52\% at the 5\% level and 88.89\% at the 10\% level. Similar patterns were observed for networks of size \( N = 500 \) and \( N = 1000 \) for \(\langle k \rangle = 4\) and \(\langle k \rangle = 10\). For \(\langle k \rangle = 20\), the proportion of significant parameters 70.37\% at the 5\% level and 86.42\% at the 10\% level for \( N = 500 \) and 69.14\% at the 5\% level and 85.19\% at the 10\% level for \( N = 1000 \). These findings suggest that the statistical significance of the core-periphery structure in the \(\mathbb{S}^1/\mathbb{H}^2\) model also decreases with increasing network density. We summarize our findings of the PSO model in Figs.~\ref{fig:p100ps0}, \ref{fig:p500pso}, and \ref{fig:p1000pso}, and for the $\mathbb{S}^1/\mathbb{H}^2$ model in Figs.~\ref{fig:p100s1}, \ref{fig:p500s1}, and \ref{fig:p1000s1}.
\end{justify}

\section*{Conclusion}

\begin{justify}

Our research highlights an important but comparatively unexplored feature of the PSO and \(\mathbb{S}^1/\mathbb{H}^2\) models: their extraordinary capacity to naturally embed the core-periphery structures within them as well as to produce highly clustered, scale-free random graphs in small worlds. Although hyperbolic models have been acknowledged in the literature as useful for representing important network features, our finding significantly improves their applicability to the modeling of real-world systems. In real systems, the core-periphery structures play a crucial role as fundamental components in the intermediate structural hierarchy of networks. We present an in-depth analysis of the dynamics of the core-periphery structure as a function of model parameters, highlighting that this structure arises naturally in hyperbolic networks due to the implicit connection rules and underlying hyperbolic geometry. These results provide new insights and inspiration for investigating and using hyperbolic network models. By highlighting the existence of core-periphery structures in these models, we pave the way to novel and highly accurate approaches to the understanding and modeling of real-world systems.

\end{justify}

\section*{Methods}

% For figure citations, please use "Fig" instead of "Figure".
\begin{justify}
This section begins with an introduction to hyperbolic network models, encompassing both the PSO model and the $\mathbb{S}^{1}/\mathbb{H}^{2}$ model. Then, we provide a concise summary of the techniques for detecting core-periphery structures in networks. Finally,  we conclude this section with a detailed description of the core-periphery structure finding algorithm utilized in our study, which includes an explanation of the core-periphery centralization concept and the statistical tests employed to assess the significance of such structure.
\end{justify}
\subsection*{Hyperbolic network models}
%\begin{Justify}
Hyperbolic geometry is a space of constant negative curvature, whereas Euclidean geometry is a flat space or a space with zero curvature. There are several models of hyperbolic space, e.g., the hyperboloid model, the Poincare disc model, the upper half-plane model, and the Klien model. Researchers commonly use the two-dimensional Poincare disc model to study the underlying hyperbolic geometry of complex networks, where each network node is represented by the polar coordinate $(r,\theta)$, here $r$ is the radial distance from the centre of the disc and $\theta$ is angular coordinate. The two-dimensional hyperbolic space $\mathbb{H}^{2}$ (or the Poincare disc) is represented by the interior of the Euclidean disc of unit radius:

\[
\mathbb{H}^{2}=\{(r,\theta)\in \mathbb{R}^{2};0\leq r<1, \theta \in [0,2\pi]\}
\]

The hyperbolic distance $d_{ij}$ in the Poincare disc between two points with polar coordinates $(r,\theta)$  and $(r^{'}, \theta^{'})$ is given by

\begin{equation}
d_{ij} = \frac{1}{\zeta}\cosh^{-1}\left(\cosh(\zeta r)\cosh(\zeta r') - \sinh(\zeta r) \sinh(\zeta r')\cos\Delta\theta\right)
\label{eq:first}
\end{equation}

Where \(\zeta=\sqrt{-K}\),  we set \(\zeta=1\); \(K\) is the curvature of the hyperbolic space, and \(\Delta\theta=\pi-|\pi-|\theta -\theta'| |\) is the angular distance between the points. Furthermore, according to Ref.\cite{krioukov2010hyperbolic}, the hyperbolic distance above in Eq. ~\eqref{eq:first} can be expressed as:

\begin{equation}
x \approx r + r^{'} + \frac{\zeta}{2} \ln\left(\frac{\Delta\theta}{2}\right)
\label{eq:second}
\end{equation}

For the \(2\sqrt{e^{-\zeta r} + e^{\zeta r{'}}} < \Delta\theta\) and the sufficiently large \(\zeta r\) and \(\zeta r{'}\).

\subsection*{The Popularity-Similarity Optimization model (PSO Model)}

The popularity-similarity optimization model \cite{papadopoulos2012popularity} is one of the hyperbolic network models which generates the complex network within the native disk representation of the hyperbolic plane described above. The basic idea is that the nodes are sequentially positioned within the disc, and the connection between them is established based on probabilities determined by their hyperbolic distances. The model has the five parameters as follows;
\begin{center}
\begin{tabular}{|p{4cm}|p{9cm}|}
\hline
Parameters & Description \\
\hline
$K<0$ & Curvature of the hyperbolic space. \\
$N \in \mathbb{N}$ & Number of nodes. \\
$m=\frac{<k>}{2} \in \mathbb{N}$ & Half of the average degree $<k>$. \\
$\beta = \frac{1}{\gamma-1} \in (0,1]$ & Popularity fading parameter where $\gamma$ is the power law coefficient. \\
$T \in [0,1]$ & Average clustering coefficient of the generated network. \\
\hline
\end{tabular}
\end{center}

\

The network construction procedure is as follows: 

\begin{itemize}
    \item[\color{black}\ding{117}] At the beginning, the network is empty and the nodes iteratively appear on the disc.
    \item[\color{black}\ding{117}] At iteration, $t=1,2, ...,N$ the new node $t$ appears with the radial coordinates as $r_{t} = \frac{2}{\zeta}\ln(t)$ and the angular coordinate $\theta_{t}$ uniformly sampled from $[0,2\pi]$.
    
    \item[\color{black}\ding{117}] All previous nodes $s<t$ increase their radial coordinates as follows $r_{s}(t) = \beta r_{s} + (1 - \beta)r_{t}$ to incorporate the popularity fading;

    \item[\color{black}\ding{117}] Furthermore, a new node $t$ is connected to the existing nodes according to the following rule: if the number of existing nodes is less than or equal to $m$, then $t$ is connected to all of them. Otherwise, if $T = 0$, then node $t$ is connected to the m closest nodes having the least hyperbolic distance $x_{st}$. For nodes with polar coordinates, $(r_{t}, \theta_{t})$ and $(r_{s}, \theta_{s})$ this distance $x_{st}$ is calculated using the hyperbolic law of cosines as defined in Eq. \eqref{eq:second}
    
    \item[\color{black}\ding{117}] for the case $T>0$ the connections of the node $t$ are established to the previous nodes  $s<t$  based on the probabilities depending upon the hyperbolic distance as follows:
    \begin{equation}
            p(x_{st}) = \frac{1}{1 + \exp({\frac{\zeta}{2T}(x_{st} - R_{t})})},
            \label{eq:third}
    \end{equation}
   Here, the distance $R_{t}$ is the current radius of the disc, which ensures that node $t$ is linked to the number of nodes $m$. It is configured as follows: \\
 For $\beta < 1$,
\begin{subequations}
    \begin{align}
        R_{t} &= r_{t} - \frac{2}{\zeta} \ln\left[\frac{2T}{\sin(T\pi)} \frac{\left(1 - e^{-\frac{\zeta}{2}(1-\beta)r_{t}}\right)}{m(1-\beta)}\right]  
        \label{eq:fourth}
    \end{align}
For $\beta = 1$, the above equation 5(a) gets reduced to the form:
\begin{align}
        R_{t} &= r_{t} - \frac{2}{\zeta} \ln\left(\frac{T}{\sin(T\pi)}\right) \frac{\zeta r_{t}}{m} 
\end{align}
    
\end{subequations}

\item[\color{black}\ding{117}] The process continues until the number of nodes $N$ has been introduced.
    
\end{itemize}

\subsection*{\(\mathbb{S}^1/\mathbb{H}^2\) Model}
\justify

\(\mathbb{S}^1\) model is quite simpler than the \(\mathbb{H}^2\) model, Here instead of radial and angular coordinates \(r_{i},\theta_{i}\), each node is represented by the hidden variable \((\kappa_{i},\theta_{i})\), where the hidden variable \(\kappa_{i}\) is the expected degree of the node \(i\), and \(\theta_{i}\) is the angular coordinates of the node \(i\) in the circle of radius \(N/2\pi\).

\(N\) nodes are initially positioned on a one-dimensional sphere (a circle) in the \(\mathbb{S}^1\) model \cite{serrano2008self}, with each node assigned a hidden variable \(\kappa_i\) in the range \([\kappa_0, \infty)\), where \(i = 1, 2, \ldots, N\) and \(\kappa_0\) minimum expected degree in the generated network. Then, pairs of nodes establish connections depending on a probability that takes into account both the hidden variables and the angular distance.

According to the procedure outlined below \cite{garcia2019mercator}, in the thermodynamic limit, \(\kappa_i\) represents the anticipated degree \(\bar{k}_i\) of node \(i\). As a result, the connection rule is simple to understand and states that nodes that are closer together in the network's hidden metric space have a higher probability of forming connections, whereas nodes with higher degrees establish longer connections. The hidden variable \(\kappa_i\) can be mapped to the radial coordinate \(r_i\) in the native representation of the hyperbolic plane \(\mathbb{H}^2\), and the hyperbolic distance between the nodes, which expresses the influence of both the node degrees and similarities, determines the connection probability.

Similarly, as in the PSO model, here we have the parameters, \(N\) The total number of nodes, \(<k>\) The average degree, \(\gamma\) The exponent of the degree distribution, following power-law: \(P(k) \sim k^{-\gamma}\). Although these models can accommodate various degree distributions, for this case, we restrict the use of power-laws with \(\gamma>2\) to generate networks with properties similar to those in the PSO model and lastly, The parameter \(\alpha\), where \(1 < \alpha\), controls the average clustering coefficient \(c\) of the resulting network (\(\lim_{\alpha\to1} c \approx 0\)).

\

The procedure for generating a \(\mathbb{S}^1\) network model, comprising \(N\) nodes, is as follows:

\begin{itemize}
    \item[\color{black}\ding{117}] For each node \(i\), an angular coordinate \(\theta_i\) is randomly  uniformly sampled from the interval \([0, 2\pi)\).

\item[\color{black}\ding{117}] For each node \(i\), a hidden variable \(\kappa_i\) is sampled from the interval \([ \kappa_0, \infty)\) according to the distribution \(\rho(\kappa) = (\gamma - 1) \cdot \kappa^{-\gamma} / \kappa_0^{-\gamma}\), where \(\kappa_0 = (\gamma - 2) / (\gamma - 1) \cdot \langle \kappa \rangle\).

\item[\color{black}\ding{117}] Each pair of nodes \(i\) and \(j\) is connected with a certain probability:

\[
p_{ij} = \frac{1}{1 + \left(\frac{N \cdot \Delta \theta_{ij}}{2\pi \cdot \mu \cdot \kappa_i \cdot \kappa_j}\right)^\alpha}
\]

Where \(\Delta\theta_{ij} = \pi - |\pi - |\theta_i - \theta_j||\) represents the angular distance between nodes, and \(\mu = \frac{\alpha}{2\pi \langle k \rangle} \cdot \sin\left(\frac{\pi}{\alpha}\right)\).

For ease of comparison with the PSO model, the hidden variable is mapped into a radial coordinate in the native representation of the hyperbolic plane (at \(K = -1\) curvature). This transformation was carried out as follows:

\begin{equation}
r_i = \hat{R} - 2 \ln \frac{k_i}{k_0},
\label{eq:fifth}
\end{equation}

Where  \(\hat{R}\) is calculated as \(2 \ln \frac{N}{\mu \pi \kappa_0^2}\). It should be noted that in this hyperbolic representation, specifically in the \(\mathbb{H}^2\) model, the connection probability Eq. ~\eqref{eq:fifth} takes the form:
\[
p_{ij} = \frac{1}{1 + e^{\frac{\alpha}{2} \cdot (x_{ij}-\hat{R})}}
\]
This connection probability depends on the hyperbolic distance \(x_{ij}\) as the connection probability in Eq. ~\eqref{eq:second}.

\section*{Core-periphery detection in networks}

We present an iterative algorithm that generates a core-periphery profile\cite{Rossa2013} to the network. This paves the way to introduce the notion of an overall network centralization index. In networks with an ideal core-periphery structure, core nodes are adjacent to core nodes, core nodes are adjacent to peripheral nodes, but peripheral nodes are not adjacent to each other. In real-world networks data there exists weak connection between peripheral nodes. 
 
Let $A=[a_{i,j}]$ be the adjacency matrix for the network $G$, where $a_{ij}> 0$ represents the weight of the edge between nodes $i$ and $j$ in an undirected connected network with nodes $N=\{1, 2, \ldots, n\}$. Let $m_{ij}$ represent the probability that a random walker, located at node $i$, transitions to node $j$, define as $m_{ij} = \frac{a_{ij}}{\sum_{h} a_{ih}}$. Furthermore, let $\pi_i > 0$ denote the asymptotic probability of being at node $i$, defined as: $\pi_i = \frac{d_i}{\sum_{i} d_i}$, where $d_i$ represents the weighted degree of node $i$. The weighted degree $d_i$ is calculated as: $d_i = \sum_j a_{ij}$.

For a subnetwork $S$ comprising nodes from the original network $N$, the persistence probability $\alpha_S$ reflects the likelihood that a random walker, currently positioned within any node of $S$, remains within $S$ during the next time step. The calculation of $\alpha_S$ can be explicitly defined as:

\[
\alpha_S = \frac{\sum_{i,j \in S} \pi_i m_{ij}}{\sum_{i \in S} \pi_i} = \frac{\sum_{i,j \in S} a_{ij}}{\sum_{i \in S} a_{ij}}.
\]

The authors \cite{Rossa2013} argue that for an ideal core-periphery (CP) structure $\alpha_S = 0$ since there would be no links between peripheral vertices. Thus, their objective is to identify the $\alpha$-periphery, which is the largest subnetwork $S$, such that $\alpha_S \leq \alpha$ for some $0 < \alpha < 1$. In other words, if a random walker is in the $\alpha$-periphery, it will exit the sub-network at the next step with probability $1 - \alpha$.

The steps for finding the core-periphery profile $\alpha_k$ of the network are outlined below.

First, we select a random node $i$ among those with the lowest weighted degree centrality. Without loss of generality, let the selected node be 1. Thus, $S_1 = \{1\}$ and hence $\alpha_1 := \alpha_{S_1} = 0$.

In step $k$, the core-periphery profile $\alpha_k$ is define as:

\[
\alpha_k = \min_{h \in N/S_{k-1}} \alpha_{S_{k-1} \cup \{h\}}.
\]

Finally, This gives us the core-periphery profile $0\leq \alpha_1 \leq \alpha_2 \leq  \ldots \leq \alpha_n = 1 $.

In Ref.\cite{Rossa2013}, a measure of the strength of the core-periphery structure called the core-periphery centralization (cp-centralization), is provided and is given by

\begin{equation}
C= 1- \frac{2}{n-2} \sum_{k=1}^{n-1} \alpha_k
\label{eq:six}
\end{equation}

Core-periphery centralization ($C$) measures the extent to which a network exhibits a core-periphery structure. A high value of $C$ indicates a clear core-periphery structure, with $C=1$ resembling a star network, while $C=0$ signifies minimal centralization akin to a complete network.

\end{itemize}

\bibliography{sample}

%\noindent LaTeX formats citations and references automatically using the bibliography records in your .bib file, which you can edit via the project menu. Use the cite command for an inline citation, e.g.  \cite{Hao:gidmaps:2014}.

%For data citations of datasets uploaded to e.g. \emph{figshare}, please use the \verb|howpublished| option in the bib entry to specify the platform and the link, as in the \verb|Hao:gidmaps:2014| example in the sample bibliography file.

\section*{Acknowledgements}

We thank the Shiv Nadar Institution of Eminence for providing the computational facilities and the necessary infrastructure needed to carry out the present research.

\section*{Author contributions statement}

P.Y. and I.A. came up with idea, design of the experiment(s) and manuscript preparation.  I.A. conducted the experiment(s). P.Y., I.A. and N.S. analysed the results.  All authors reviewed the manuscript. 

%\section*{Additional information}

%To include, in this order: \textbf{Accession codes} (where applicable); 

\section*{Data availability:} 
Code for generating and analyzing the datasets during the current study is available in the GitHub repository: \url{https://github.com/Imran10896/cp_structure_in_hyperbolic_networks.git}.

\section*{\textbf{Competing interests}} The authors declare no competing financial interests.

%Figures and tables can be referenced in LaTeX using the ref command, e.g. Figure \ref{fig:stream} and Table \ref{tab:example}.

%\begin{figure}[htbp]
%    \centering
%    \includegraphics[width=1.0\textwidth]{PSO_size.png}
%    \caption{Size dependence of the core-periphery centralization in the PSO model. We plotted the highest weighted core-periphery centralization $C$, as a function of the number of nodes $N$. The expected average degree $\langle k \rangle$ was set to 10 in each case. The further model parameters, such as the $\beta$, and $T$ values appear as panel titles and legends. Each data point was obtained by averaging over 10 networks of the given parameter set. The error bars indicate the standard deviations among the 10 networks.}
%    \label{fig:heatmaps}
%\end{figure}

%\begin{figure}[htbp]
%    \centering
%    \includegraphics[width=1.0\textwidth]{S1_size_output.png}
%    \caption{Size dependence of the core-periphery centralization in the \(\mathbb{S}^1/\mathbb{H}^2\) model. We plotted the highest weighted core-periphery centralization $C$, as a function of the number of nodes $N$. The expected average degree $\langle k \rangle$ was set to 10 in each case. The further model parameters, such as the $\gamma$, and $\alpha$ values appear as panel titles and legends. Each data point was obtained by averaging over 10 networks of the given parameter set. The error bars indicate the standard deviations among the 10 networks.}
 %   \label{fig:heatmaps}
%\end{figure}

\begin{figure}[htbp]
    \centering

    \includegraphics[width=1.0\textwidth]{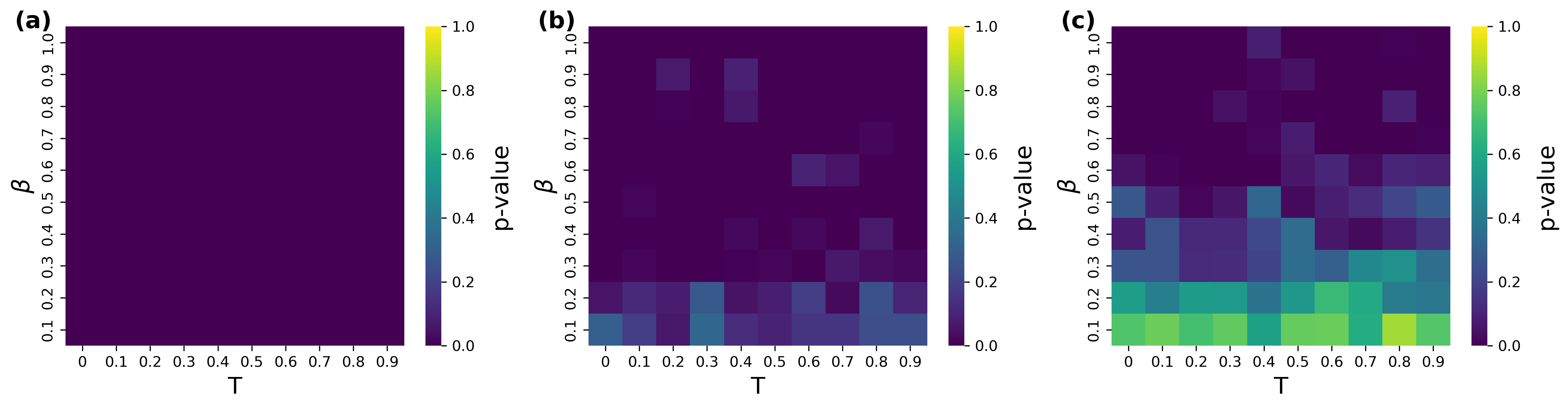}
    \caption{$p$-value in the PSO model for the model parameters $T$ and $\beta$ for networks of size $N = 100$ and the expected average degree: (a) $\langle k \rangle = 4$, (b) $\langle k \rangle = 10$, and (c) $\langle k \rangle = 20$.}
    \label{fig:p100ps0}
\end{figure}

\begin{figure}[htbp]
    \centering
    \includegraphics[width=1.0\textwidth]{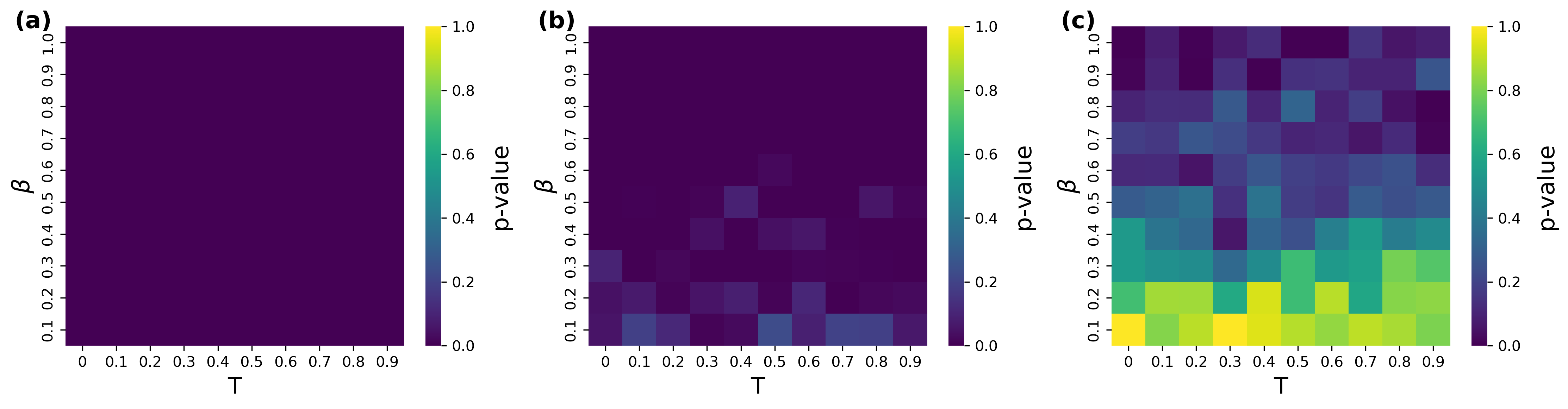}
    \caption{$p$-value in the PSO model for the model parameters $T$ and $\beta$ for networks of size $N = 500$ and the expected average degree: (a) $\langle k \rangle = 4$, (b) $\langle k \rangle = 10$, and (c) $\langle k \rangle = 20$.}
    \label{fig:p500pso}
\end{figure}

\begin{figure}[htbp]
    \centering
    \includegraphics[width=1.0\textwidth]{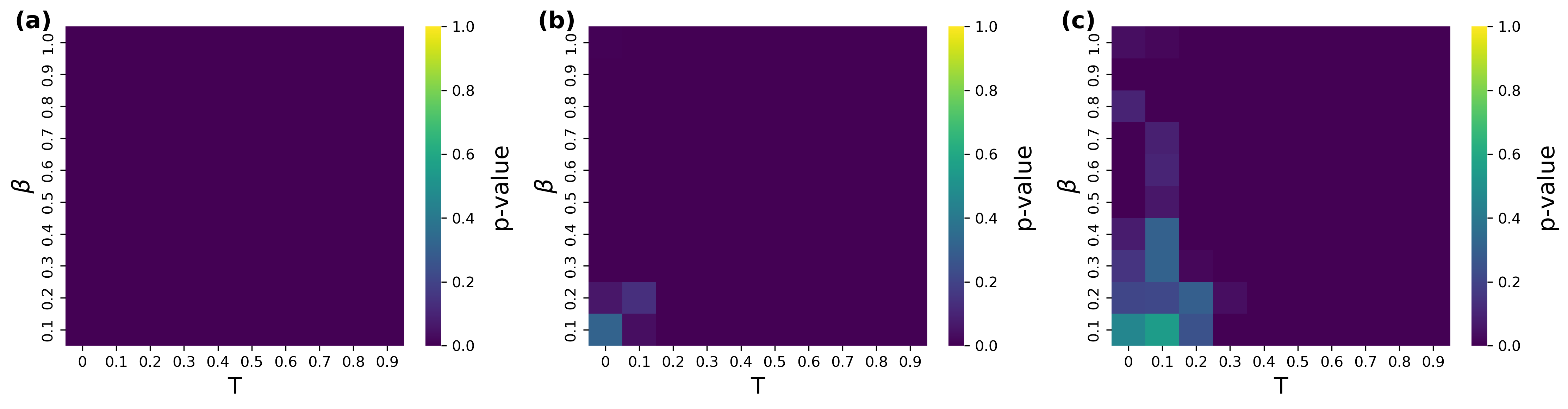}
    \caption{$p$-value in the PSO model for the model parameters $T$ and $\beta$ for networks of size $N = 1000$ and the expected average degree: (a) $\langle k \rangle = 4$, (b) $\langle k \rangle = 10$, and (c) $\langle k \rangle = 20$.}
    \label{fig:p1000pso}
\end{figure}

\begin{figure}[htbp]
    \centering
    \includegraphics[width=1.0\textwidth]{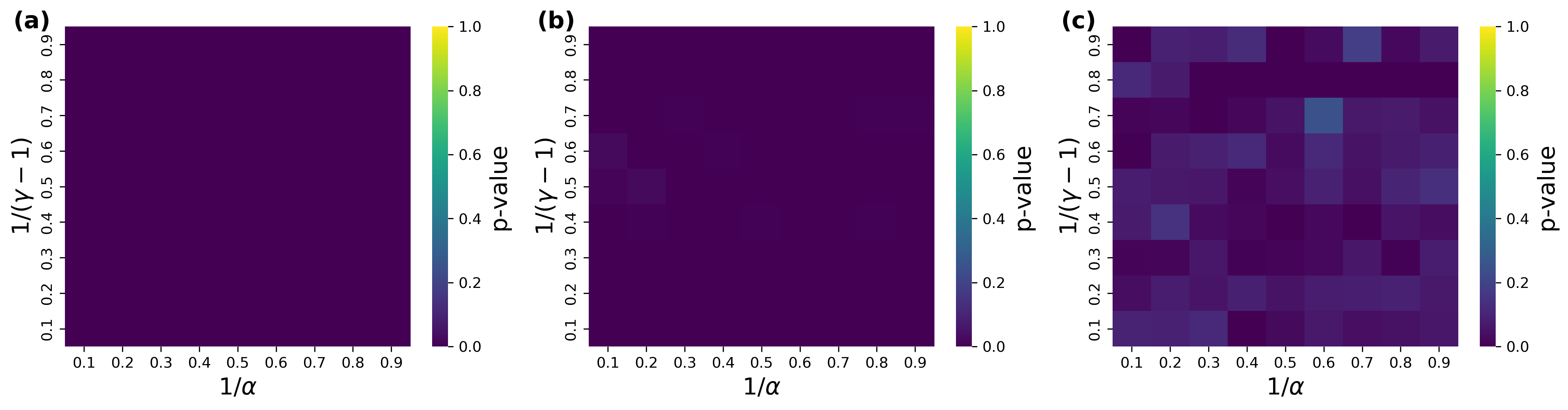}
   \caption{$p$-value in the $\mathbb{S}^1/\mathbb{H}^2$ model for the model parameters $1/\alpha$ and $1/(\gamma-1)$ for networks of size $N = 100$ and the expected average degree: (a) $\langle k \rangle = 4$, (b) $\langle k \rangle = 10$, and (c) $\langle k \rangle = 20$.}
    \label{fig:p100s1}
\end{figure}

\begin{figure}[htbp]
    \centering
    \includegraphics[width=1.0\textwidth]{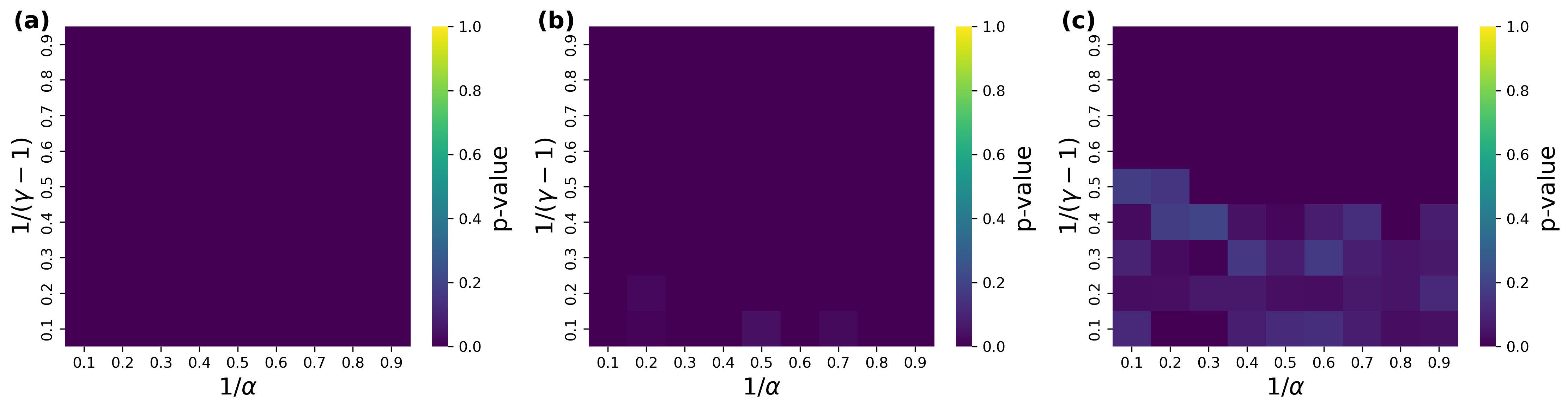}
    \caption{$p$-value in the $\mathbb{S}^1/\mathbb{H}^2$ model for the model parameters $1/\alpha$ and $1/(\gamma-1)$ for networks of size $N = 500$ and the expected average degree: (a) $\langle k \rangle = 4$, (b) $\langle k \rangle = 10$, and (c) $\langle k \rangle = 20$.}
    \label{fig:p500s1}
\end{figure}

\begin{figure}[htbp]
    \centering
    \includegraphics[width=1.0\textwidth]{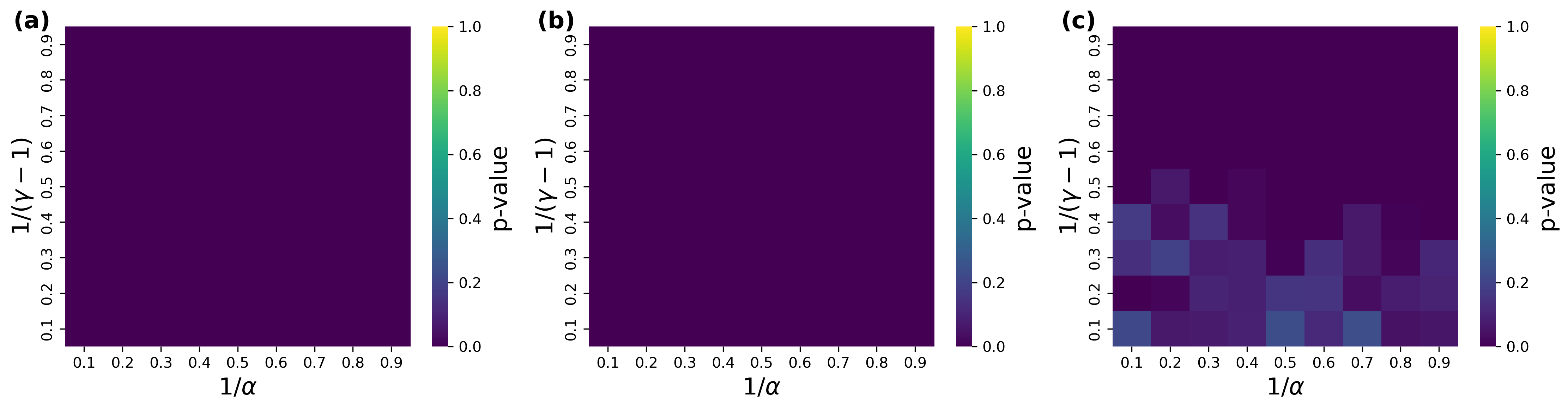}
    \caption{$p$-value in the $\mathbb{S}^1/\mathbb{H}^2$ model for the model parameters $1/\alpha$ and $1/(\gamma-1)$ for networks of size $N = 1000$ and the expected average degree: (a) $\langle k \rangle = 4$, (b) $\langle k \rangle = 10$, and (c) $\langle k \rangle = 20$.}
    \label{fig:p1000s1}
\end{figure}

\end{document}